\documentclass[aps,groupedaddress,nofootinbib]{revtex4}

\usepackage[latin1]{inputenc}
\usepackage{graphicx}
\usepackage{amssymb,amsmath}

\begin{document}

\preprint[\leftline{KCL-PH-TH-2012-31, LCTS/2012-15, CERN-PH-TH-2012-207}

\title{Asymptotic Analysis of the Boltzmann Equation for Dark Matter Relics
in the presence of a Running Dilaton and Space-Time Defects}

\author{Carl M. Bender\footnote{Permanent address: Department of Physics,
Washington University, St.~Louis, MO 63130, USA.}}
\email[]{cmb@wustl.edu}
\affiliation{Department of Physics, King's College London, Strand, London WC2R
2LS, UK}

\author{Nick E. Mavromatos\footnote{Currently also at: CERN, Physics Department,
Theory Division, CH 1211 Geneva 23, Switzerland.}}
\email[]{Nikolaos.Mavromatos@cern.ch}
\affiliation{Department of Physics, King's College London, Strand, London WC2R
2LS, UK}

\author{Sarben Sarkar}
\email[]{sarben.sarkar@kcl.ac.uk}
\affiliation{Department of Physics, King's College London, Strand, London WC2R
2LS, UK}

\date{\today}

\begin{abstract}
The interplay of dilatonic effects in dilaton cosmology and stochastic quantum
space-time defects within the framework of string/brane cosmologies is examined.
The Boltzmann equation describes the physics of thermal dark-matter-relic
abundances in the presence of rolling dilatons. These dilatons affect the
coupling of stringy matter to D-particle defects, which are generic in string
theory. This coupling leads to an additional source term in the Boltzmann
equation. The techniques of asymptotic matching and boundary-layer theory, which
were recently applied by two of the authors (CMB and SS) to a Boltzmann
equation, are used here to find the detailed asymptotic relic abundances for all
ranges of the expectation value of the dilaton field. The phenomenological
implications for the search of supersymmetric dark matter in current colliders,
such as the LHC, are discussed.
\end{abstract}

\maketitle

\section{Introduction}
\label{s1}
To evaluate candidates for cold dark matter (DM) it is necessary to compute
relic abundances including physics beyond the standard model. String-theory
considerations provide a natural source of such physics.

Recently, a {\it global} asymptotic analysis was performed on the (Riccati-type)
Boltzmann differential equation that describes the evolution of the thermal DM
relic abundances in an expanding universe \cite{R1}. It was shown that
boundary-layer theory, which makes use of asymptotic matching \cite{R2}, can
give a consistent approximate solution to this Riccati equation in two
physically interesting cases: (i) standard Friedman-Robertson-Walker (FRW)
cosmology \cite{R3}, and (ii) dilatonic string cosmology \cite{R4,R5}. In case
(i) the freeze-out and post-freeze-out {\it regions} (we emphasize that
these are regions and not isolated points) for the DM abundances were defined
using this novel approach. In case (ii) the Boltzman equation of case (i) is
modified by the addition of a rolling dilaton source term derivable from string
theory and proportional to the dilaton cosmic rate $\frac{d\Phi}{dt}$. The
effects of the rolling dilaton on cold DM abundances were calculated and it was
shown that there is a large-time power-law decay of the DM abundance (with
calculable corrections). The latter results explain the findings of \cite{R6} on
the dilution of DM relic abundances in the current epoch in supersymmetric
theories with rolling dilatons. This dilution may significantly affect the
available parameter space (after the appropriate cosmological constraints from
WMAP \cite{wmap} are taken into account) and, in turn, may affect the searches
for supersymmetry at colliders such as the LHC \cite{R7}.

The analysis cited above does not include the effect of a cosmological
background due to effectively point-like defects (quantum space-time foam),
which are generically found in models based on string theory \cite{R8}. Dilatons
are coupled to the foam through the string coupling constant. This foam modifies
the effect of the dilaton in the evolution of DM and can even \emph{dominate}
asymptotically in the absence of dilaton effects. In our model of space-time
foam the universe is represented as a brane, with three large spatial
longitudinal dimensions, embedded in a higher-dimensional bulk space. The
``foamy'' structures are provided by stringy membrane (D-brane) defects,
compactified appropriately along extra-dimensional manifolds. From the point of
view of a four-dimensional observer the defects appear to be point-like
(D particles).\footnote{D particles are in the spectrum of some but not all
string theories; that is, they exist in the spectrum of type IIA but not type
IIB string theory. Even when D particles are not in the spectrum, compactified
higher dimensional branes may seem like effective D particles for observers on
the brane world.} As the D-brane world moves in the bulk, the D particles cross
it and thus appear to the four-dimensional observer as stochastic space-time
structures, flashing on and off. The stochasticity in target-space is attributed
to quantum fluctuations of the D particles, viewed as stringy dynamical entities
embedded in the bulk space. The dilaton $\Phi$ directly affects this process
because its vacuum expectation value determines the string coupling $g_s$. This
paper investigates the interplay between the dilaton and a background of
space-time defects and their effects on the asymptotic behavior of relic
abundances.

In Sec.~\ref{sec:2} we review the main results of \cite{R1} concerning the DM
relic density for asymptotically long times in standard and dilatonic
cosmologies. This serves to introduce the powerful technique of asymptotic
matching \cite{R2} used in Ref.~\cite{R1}. In Sec.~\ref{sec:3} we apply the
analytical methods of \cite{R1} to the case of D-particle stochastic foam (D
foam) in the presence of relaxing dilatons. The D foam is characterized by a
source that differs from the source in dilatonic cosmology. Various classes of
asymptotic behaviors are determined by the expectation value of $\Phi$. In
Sec.~\ref{sec:4} we discuss the phenomenology of these models. We address the
combined effects of the running-dilaton and D-foam sources on the thermal DM
relic abundances today and the associated constraints implied by the current LHC
phenomenology. Finally, a technical discussion of the thermodynamic properties
of the various types of universes in the presence of sources, which are examined
in this article, is given in the Appendix. There, we explain in detail how it is
possible to define an entropy function that is conserved in the presence of
nontrivial source terms in the Boltzmann equation; this function allows a
thermodynamic interpretation of the respective cosmological equations.

\section{Review of asymptotic solutions to the Boltzmann equations for relic
abundances in standard and dilatonic cosmologies
\label{sec:2}}

For a DM species $X$ of mass $m_X$ the evolution of $Y(x)\equiv\mathcal{N}/s$,
the number density $\mathcal{N}$ per entropy density $s$, is governed by the
Boltzmann equation \cite{R3}
\begin{equation}
Y'(x)=-\lambda x^{-n-2}\left[Y^2(x)-Y_{{\rm eq}}^2(x)\right],
\label{e1}
\end{equation}
where $x\equiv m_X/T$ is the dimensionless independent variable and $T$ is the
temperature. This Riccati equation does not include any dilatonic effects of
string theory. We are primarily interested in epochs of the universe for which
$m_X>T>T_0$, where $T_0$ is the current temperature of the universe. The
integer $n=0,\,1,\,2,\,\ldots$ comes from a partial-wave analysis of the
scattering of DM particles: $n=0$ refers to $s$-wave scattering, $n=1$
characterizes $p$-wave scattering, and so on. The parameter $\lambda$ is a
dimensionless measure of the scattering of DM particles and is regarded as a
large number ($\lambda\gg1$). If we parametrize the thermally averaged annihilation
cross-section $\langle\sigma v\rangle=\sigma_0\,x^{-n}$ with $n=0,\,1,\,\dots$
for $(s,\,p,\,\dots)$-wave DM annihilation, and the Hubble parameter as $H=H_m\,
x^{-2}$, then $\lambda\equiv\sigma_0\,m_X^3/H_m$ \cite{R3}. For bosonic remnants
the function $Y_{\rm eq}(x)$ is the distribution \cite{R10}
\begin{equation}
Y_{\rm eq}(x)=A\int_0^\infty ds\frac{s^2}{e^{\sqrt{s^2+x^2}}-1},
\label{e2}
\end{equation}
where $A=0.145g/g_*$, $g$ is the degeneracy factor for the DM species, and $g_*$
counts the total number of massless degrees of freedom \cite{R3}.

A closed-form analytical solution to the Riccati equation (\ref{e1}) is
unavailable, so an approximate heuristic approach is customarily used to treat
this equation: As the universe cools and $x$ increases, the nature of the
solution $Y(x)$ to (\ref{e1}) changes rapidly in the vicinity of a value $x=
x_f$, the so-called \textit{freeze-out} point, and as $x\to\infty$ the solution
$Y(x)$ approaches the constant $Y_\infty$, called the \textit{relic abundance}.
One approximation is made for $x<x_f$ and another is made for $x>x_f$. The
solutions in the two regions are then patched at $x=x_f$. The value $x_f$ is
determined from equating the interaction rate of the DM particle and the
expansion rate of the universe, a sensible physical criterion.

This approach gives a reasonably accurate determination of $Y_\infty$ and, prior
to the work of \cite{R1}, it has been widely adopted \cite{R3}. However, this
splitting into two regions is only a pragmatic convenience and there is really
no precise value $x_f$. Rather, there may be (in a sense to be specified) a
freeze-out \emph{region}. Because the differential equation (\ref{e1}) is
first order, its solution is completely determined by \emph{one} initial
condition, namely $Y(0)$. The usual method of splitting (\ref{e1}) into two
approximate first-order equations, which are valid in each of two regions,
requires two conditions, an initial condition and a patching condition. The
value of $x_f$ becomes explicitly involved in the determination of
$Y_\infty$ even though the mathematical theory of differential equations does
not require this. To avoid this unsatisfactory mathematical treatment (which is
common in the literature), two of the current authors (CMB and SS) presented in
\cite{R1} a detailed analysis of the associated Riccati equations using applied
mathematical methods commonly used in fluid mechanics. A key concept is that the
freeze-out region can, at least in physically relevant cases, be considered as a
boundary layer. The solutions in the two regions can then be matched
asymptotically. Before reviewing the solution of (\ref{e1}) for large $x$ we
introduce the Boltzmann equation in the presence of a dilaton background.

In the case of rolling dilaton cosmologies \cite{R4} the thermal DM relic
abundance is characterized by the presence of a linear \emph{sink} term, which
is proportional to the rate of the rolling dilaton field $\frac{d\Phi}{dt}$
\cite{R5}. In theories with scale-factor duality \cite{R4}, we have
\begin{equation}
\Phi(t)=\Phi_0\,{\rm log}\,a(t),
\label{e3}
\end{equation}
where $a(t)$ is the scale factor of the expanding universe. In eras where the
temperature $T$ satisfies $m_X>T>T_0$,
\begin{equation}
Y'(x)=-\lambda x^{-n-2}\left[Y^2(x)-Y_{\rm eq}^2(x)\right]+\Phi_0Y(x)/x.
\label{e4}
\end{equation}
Here, $\Phi_{0}$ is a \emph{negative} dimensionless constant of order 1 that
appears in the general expression for the dilaton field as a function of cosmic
time $t$. For $\Phi_0=-\phi<0$, the string coupling $g_s=e^\Phi$ becomes
\emph{perturbatively} small for large times and vanishes asymptotically as $t\to
\infty$. Thus, the $\sigma$-model perturbative picture suffices to describe the
features of cosmology at large-times. As shown in \cite{R1}, the presence of the
particular dilaton source in (\ref{e4}) gives a solution for $Y(x)$ whose
behavior is \emph{qualitatively different} from the solution for $Y(x)$ in
(\ref{e1}).

\subsection{Boundary-Layer Theory}

Since $\lambda$ is large, the highest derivative in both equations (\ref{e1})
and (\ref{e4}) is multiplied by a small parameter, which implies that these
equations may be treated by using boundary-layer techniques \cite{R2} and leads
to the concept of a freeze-out \emph{region} as opposed to a freeze-out point
\cite{R1}. We rewrite (\ref{e4}) as
\begin{equation}
\frac{1}{\lambda}Z'(x)=-x^{-n-2}\left[x^{-\phi}Z^2(x)-x^\phi Y_{\rm eq}^2(x)
\right],
\label{e5}
\end{equation}
where
\begin{equation}
Z(x)\equiv Y(x)x^\phi.
\label{e6}
\end{equation}
The coefficient $1/\lambda$ of the highest-derivative term is very small. The
number of terms on the right side has been reduced from three to two; this
facilitates asymptotic matching. Outside a boundary layer (the outer region),
$Z(x)$ varies slowly. Inside a boundary layer, $Z(x)$ varies rapidly.

We have two outer regions where $Z(x)=Z^{(1)}(x)$ and $Z(x)=Z^{(2)}(x)$,
respectively. In the left outer region $Z^{(1)}(x)\approx Z_{\rm eq}(x)\equiv
x^\phi Y_{\rm eq}(x)$. To be precise, we write
\begin{equation}
Z^{(1)}(x)\sim\sum_{k=0}^\infty\lambda^{-k}Z_k^{(1)}(x).
\label{e7}
\end{equation}
On substituting $Z^{(1)}(x)$ into (\ref{e5}), we find that
\begin{eqnarray}
Z_0^{(1)}(x) &=& Ae^{-x}x^{\varphi+3/2},\nonumber\\
Z_1^{(1)}(x) &=& x^{\varphi+n+2}/2,
\label{e8}
\end{eqnarray}
and so on. The entity $x_f$ is defined to be the value of $x$ for which
\begin{equation}
Z_0^{(1)}(x)=Z_1^{(1)}(x)
\label{e9}
\end{equation}
and is a measure of where the equilibrium region ends. Equation (\ref{e9})
implies that
\begin{equation}
x_f\sim\log(2A\lambda)-(n+1/2)\log\left(x_f\right).
\label{e10}
\end{equation}
This analysis is somewhat simplified (see~\cite{R1}). The higher order terms in (\ref{e7}) are not negligible, but they lead to a series with alternating signs that is Borel summable. The Borel sum of the series leads to a multiplicative renormalization of $A$ by a factor close to $1$. In order to keep the notation simple we have not distinguished $A$ from the renomalized $A$.
Solving the equation obtained by replacing in (\ref{e10}) the symbol $\sim$ by
the equality sign gives $x_f$:
\begin{equation}
x_f=(n+1/2)W\left[\frac{(2\lambda A)^{n+1/2}}{n+1/2}\right],
\label{e11}
\end{equation}
where $W(z)$ is a Lambert function \cite{R11}. Hence the asymptotic behavior is
fully determined in terms of constants occurring in the Boltzmann equation.

The value $x_{f}$ lies in the transition \emph{region} from equilibrium to
freeze-out which is interpreted as a boundary layer. This interpretation can be
justified by the method of asymptotic matching. We define an inner variable $X$
as follows:
\begin{equation}
x=x_f+\kappa X.
\label{e12}
\end{equation}
Then $|X|$ can be large compared to $1$ but small compared to $\lambda$. Now,
for $\mathcal{Z}(X)$ we have
\begin{equation}
\frac{1}{\kappa}\mathcal{Z}'(X)=-\lambda x_f^{-n-2-\phi}\left[\mathcal{Z}^2(X)-
A^2x_f^{3+2\phi}e^{-2x_f}\right]\approx-\lambda x_f^{-n-2-\phi}\mathcal{Z}^2(X).
\label{e13}
\end{equation}
The exponential term is negligible because $x_f\approx25$ for typical values
$\lambda\approx10^{14}$ and $A\approx0.00145$.

From the principle of dominant balance \cite{R2} we have
\begin{equation}
\kappa=x_f^{n+2+\phi}/\lambda.
\label{e14}
\end{equation}
The solution to (\ref{e13}) is
\begin{equation}
\mathcal{Z}(X)=1/(X+D),
\label{e15}
\end{equation}
where $D$ is a constant of integration. This is the solution in the
boundary-layer (or freeze-out) region.

To the right of this boundary layer there is a second outer region. For large
$x$ in this region
\begin{equation}
Z'(x)\sim-\lambda x^{-n-2-\phi}Z^2(x)\qquad(x\gg1),
\label{e16}
\end{equation}
whose solution is
\begin{equation}
Z^{\rm post-freeze-out}(x)\sim\frac{1}{1/C-\lambda x^{-n-1-\phi}/(n+1+\phi)},
\label{e17}
\end{equation}
where $C$ is an integration constant.

The behaviors in the equilibrium outer region, the boundary-layer region and the
post-freeze-out outer region must be asymptotically matched. This matching
determines the constants of integration $C$ and $D$. We first match the solution
in the equilibrium region to the boundary-layer solution:
$$Z^{{\rm thermal-equilibrium}}(x)\sim2Ax^{3/2+\phi}e^{-x}\sim2A\left(x_f+
\kappa X\right)^{3/2+\phi}e^{-x_f}e^{-\kappa X}.$$
The factor of $2$ is included because two lowest-order terms of the expansion in
(\ref{e7}) are considered. Noting that $\kappa$ and $X/x_f$ are small, we get
\begin{equation}
Z^{\rm thermal-equilibrium}(x)\sim\frac{x_f^{n+2+\phi}}{\lambda(1+\kappa X)}\sim
\frac{1}{X+\lambda x_f^{-n-2-\phi}}
\label{e18}
\end{equation}
on using (\ref{e14}). Hence, comparing with (\ref{e15}), we deduce that
\begin{equation}
D=\lambda x_f^{-n-2-\phi}.
\label{e19}
\end{equation}

Similarly, (\ref{e17}) leads to
$$Z^{\rm post-freeze-out}(x)\sim\frac{1}{\frac{1}{C}-\frac{\lambda}{n+1+\phi}
(x_f+\kappa X)^{-n-1-\phi}},$$
from which we deduce that
\begin{equation}
Z^{\rm post-freeze-out}(x)\sim\frac{1}{\frac{1}{C}-\frac{\lambda}{(n+1+\phi)
x_f^{n+1+\phi}}+X}.
\label{e20}
\end{equation}
Comparing with (\ref{e15}), we get
\begin{equation}
D=\frac{1}{C}-\frac{\lambda}{\left(n+1+\phi\right)x_f^{n+1+\phi}}.
\label{e21}
\end{equation}
Finally, from (\ref{e19}) we deduce that
\begin{equation}
C=\frac{(n+1+\phi)x_f^{n+2+\phi}}{\lambda\left(n+1+\phi+x_f\right)}.
\label{e22}
\end{equation}
The leading behavior for large $x$ in the post-freeze-out region is
\begin{equation}
\label{e23}
Y(x)\sim\frac{(n+1+\phi)x_f^{n+2+\phi}}{\lambda(n+1+\phi+x_f)}x^{-\phi}.
\end{equation}
We denote the solution to (\ref{e1}) as $Y_{ns}(x)$, where {\it ns} stands for
{\it no source}. Its asymptotic value for large $x$ is obtained from (\ref{e23})
by setting $\phi=0$. The specific solution for $x_f$ in (\ref{e11}) is denoted
by $x_{f,ns}$.

The above calculation forms the basis of the following analysis that will be
given for various parameter ranges and sources in the Boltzmann equation.

\section{DM Relic Abundances: the case of a Stochastic Stringy Space-Time Foam}
\label{sec:3}

The background of stochastic D-particle foam leads \cite{R8} to the inclusion of
a positive source $\Gamma$ (as opposed to the sink in dilaton cosmology) in the
standard Boltzmann equation for the thermal relic abundance of the DM species
$X$ of mass $m_X$. In terms of the number density $\mathcal{N}$ it was shown in
Ref.~\cite{R8} that the Boltzmann equation reads
\begin{equation}
\frac{d\mathcal{N}}{dt}+3H\mathcal{N}=\Gamma(t)\mathcal{N}+C[f],
\label{e24}
\end{equation}
where $C[f]$ denotes the Boltzmann interaction terms and
\begin{equation}
\Gamma(t)=2Hm_Xa^4(t)\frac{g_s^2}{M_s^2}T(9+2m_X/T)\,\ll\Delta^2\gg,
\label{e25}
\end{equation}
where $M_s$ is the string mass scale. (The mass of a D-particle defect in the
foam is $M_s/g_s$ \cite{R8}.) The quantity $\ll\Delta^2\gg$ is a dimensionless
variable, which expresses the variance in the recoil velocities of the
D-particle defects in the foam, during their collisions with the DM particles
\cite{R8}.

The symbol $\ll\dots\gg$ denotes the average over the population of D particles
on the three-dimensional-space brane world in a given epoch of the universe.
The no-force (dust-like) behavior of the D particles, implies the following scaling of
$\ll\Delta^2\gg$ with the scale factor $a(t)$ of the four-dimensional (brane)
universe:
\begin{equation}
\ll\Delta^2\gg=\langle\Delta^2\rangle_0\,a^{-3}(t)=\langle\Delta^2\rangle_0\,
C_0^{-3}\,T^3=\langle\Delta^2\rangle_0\,m_X^3\,C_0^{-3}\,x^{-3}.
\label{e26}
\end{equation}
Here $C_0=a(t_0)T_0$ is a dimensionful constant that appears in the cooling law
of the universe; that is,
\begin{equation}
a(t)=C_0/T=a(t_0)/(1+z),
\label{e27}
\end{equation}
where $z$ is the redshift parameter. The values $z=0$, $t=t_0$, and $T=T_0$ are
correspond to the current era. This source is positive (in contrast to the sink
of dilaton cosmology) and is discussed in a more general framework in the
Appendix.

We now discuss the collision term
$\langle\sigma v\rangle\left[\left(\mathcal{N}^{(0)}\right)^2-\mathcal{N}^2
\right]$
in (\ref{e24}), where
$\mathcal{N}^{(0)}$ is the equilibrium
value of the DM number density. Eq.~(\ref{e24}) now becomes
\begin{equation}
Y'(x)=-\lambda x^{-n-2}\left[Y^2(x)-Y_{\rm eq}^2(x)\right]+
\frac{2C_0^4 g_s^2}{m_X^2M_s^2}\ll\Delta^2\gg x^2(9+2x)\,Y(x).
\label{e28}
\end{equation}
Hence, the Boltzmann equation (\ref{e28}) becomes
\begin{equation}
Y'(x)=-\lambda x^{-n-2}\left[Y^2(x)-Y_{\rm eq}^2(x)\right]+g_s^2\,\frac{2C_0\,
m_X}{M_s^2}\,\langle\Delta^2\rangle_0\,\left(9+2x\right)\,Y(x)/x.
\label{e29}
\end{equation}

There is an implicit dilaton dependence in (\ref{e29}) that needs to be made
explicit. The string coupling $g_s$ is the exponential of the dilaton, $g_s
=g_0\,\exp\left(\left\langle\Phi\right\rangle\right)$, and so
\begin{equation}
g_s\sim g_0\,a^{-\phi}=g_0\left(C_0/m_X\right)^{-\phi}x^{-\phi}.
\label{e30}
\end{equation}
Hence, for consistency we must incorporate \emph{both} the dilaton sink and the
source induced by D-particle foam in the Boltzmann equation in a combined
source. The resulting Boltzmann equation is
\begin{equation}
Y'(x)=-\lambda x^{-n-2}\left[Y^2(x)-Y_{\rm eq}^2(x)\right]-
\mathcal{S}(x,\phi)Y(x)/x,
\label{e31}
\end{equation}
where
\begin{equation}
S(x,\phi)=\phi-\zeta_\phi(9+2x)x^{-2\phi}
\label{e32}
\end{equation}
and
\begin{equation}
\zeta_\phi\equiv2g_0^2 C_0^{1-2\phi}m_X^{1+2\phi}\langle\Delta^2\rangle_0/M_s^2.
\label{e33}
\end{equation}

We are especially interested in the regime of temperatures $m_X\gg T$; that is,
as $x\to\infty$. However, the asymptotic matching requires a knowledge of the
solution for higher $T$ as well. From (\ref{e32}) it is clear that for the case
$\phi=1/2$ the $x$-dependence of the source and sink coincide for large $x$.
Just as in (\ref{e5}), it is convenient to rewrite (\ref{e31}) in the form of a
differential equation with the nonlinear terms on the right side. We introduce
the function
\begin{equation}
g(x,\phi)\equiv x^\phi\exp\left[\zeta_\phi\left(\frac{9}{2\phi}-\frac{2x}{1-2
\phi}\right)x^{-2\phi}+\frac{2\zeta_{\phi}}{1-2\phi}\right],
\label{e34}
\end{equation}
which is smooth at $\phi=1/2$, and the function
\begin{equation}
\label{e35}
f(x,\phi)\equiv x^{-n-2}/g(x,\phi).
\end{equation}
We then define
$$Z(x,\phi)\equiv g(x,\phi)Y(x),$$
which is the analog of (\ref{e6}), and
$$Z_{\rm eq}(x,\phi)\equiv g(x,\phi)Y_{\rm eq}(x).$$
The Riccati equation satisfied by $Z(x,\phi)$ is
\begin{equation}
\frac{dZ(x,\phi)}{dx}=-\lambda f(x,\phi)\left[Z^2(x,\phi)-Z_{\rm eq}^2(x,\phi)
\right],
\label{e36}
\end{equation}
which is similar in structure to (\ref{e5}). The explicit form of $f(x,\phi)$ is
\begin{equation}
f(x,\phi)=x^{-n-2-\phi}\exp\left[-\zeta_{\phi}\left(\frac{9}{2\phi}-\frac{2x}
{1-2\phi}\right)x^{-2\phi}-\frac{2\zeta_\phi}{1-2\phi}\right].
\label{e37}
\end{equation}

The dominant asymptotic behavior of $f(x,\phi)$ as $x\to\infty$ changes
according to the value of $\phi$; $f(x,\phi)$ decays for $\phi>1/2$ and $f(x,
\phi)$ increases exponentially for large $x$ for $\phi<1/2$. Note that the
phenomenologically relevant quantity is the Hubble-constant-free-relic
abundance, $\Omega\,h^2=m\mathcal{N}/\rho_0^c$, where $\rho_0^c$ is the critical
density today and $\mathcal{N}$ is the number density of the DM species. This is
the quantity that is measured in experiments. For DM species $X$ with mass $m_X$
it is given by \cite{R3}
\begin{equation}
\Omega_X h^2=m_X^4Y(x)/x^3.
\label{e38}
\end{equation}

The modification of $\Omega_X$ can be compared to the standard (source-free)
relic density by considering the phenomenologically interesting ratio
\begin{equation}
\mathcal{R}\equiv\lim_{x\to\infty}\frac{\Omega_X}{\Omega_X^{\rm source-free}}
\sim\lim_{x\to\infty}\frac{Y(x)}{Y_{ns}(x)},
\label{e39}
\end{equation}
where $\Omega_X^{\rm source-free}$ denotes the relic density of the DM species
$X$ in the standard cosmology case with constant dilaton and no space-time
foam. We now systematically consider the behavior of the solution to the
Boltzmann equation for various values of $\phi$.

\subsection{The case of $\phi$ near $1/2$}

To investigate the behavior near $\phi=1/2$ we let $\phi=1/2-\delta$ and treat
$\delta$ as small. We write $Z_\delta(x)\equiv Z(x,1/2-\delta)$. The Riccati
equation satisfied by $Z_\delta(x)$ in (\ref{e36}) is
\begin{equation}
Z_\delta'(x)=-\lambda f_\delta(x)\left[Z_\delta^2(x)-Z_{eq,\delta}^2(x)\right],
\label{e40}
\end{equation}
where $Z_{eq,\delta}(x)\equiv Z_{\rm eq}(x,1/2-\delta)$ for small $\delta$ and
\begin{equation}
f_\delta(x)\approx x^{-n-5/2+\delta+2\eta_\delta}\exp\left(-9\eta_\delta/x
\right)
\label{e41}
\end{equation}
with $\eta_\delta\equiv\zeta_{1/2-\delta}$. Moreover, we have
\begin{equation}
Z_{eq,\delta}(x)\sim Ax^{2-\delta-2\eta_\delta}e^{-x}.
\label{e42}
\end{equation}

As we did in Sec.~\ref{sec:2}, we argue that for large $x$ in the
post-freeze-out outer region, $Z_\delta(x)\approx Z_\delta^{\rm post-freeze-out}
(x)$, where
\begin{equation}
\frac{d}{dx}Z_\delta^{\rm post-freeze-out}(x)=-\lambda f_\delta(x)\left[
Z_\delta^{\rm post-freeze-out}(x)\right]^2.
\label{e43}
\end{equation}
The solution to (\ref{e43}) is
\begin{equation}
Z_\delta=\frac{1}{\mathcal{C}_\delta^{-1}-\lambda\frac{x^{-n-3/2+\delta+
2\eta_\delta}}{n+3/2-\delta-2\eta_\delta}},
\label{e44}
\end{equation}
and $\mathcal{C}_\delta$ is an integration constant to be determined.

In the equilibrium outer region, following (\ref{e5}), (\ref{e7}), and
(\ref{e9}), we substitute
\begin{equation}
Z_\delta(x)\sim\sum_{k=0}^\infty\lambda^{-k}Z_{k,\delta}(x)
\label{e45}
\end{equation}
into (\ref{e40}). This leads to
$Z_{0,\delta}(x)=Z_{eq,\delta}(x)$
and
$$Z_{1,\delta}(x)=-\frac{1}{2f_\delta(x)}\frac{d}{dx}\log Z_{eq,\delta}.$$
The value $x=x_f$, which characterises the freeze-out region, is determined by
$Z_{0,\delta}(x)=Z_{1,\delta}(x),$
and we again obtain (\ref{e10}).

In the inner (freeze-out) region we introduce $X$ as in (\ref{e12}). The
resulting equation for $\mathcal{Z_\delta}(X)$ is
\begin{equation}
\frac{1}{\kappa}\frac{d}{dX}\mathcal{Z_\delta}\left(X\right)=-\lambda x_f^{-n-
5/2+\delta+2\eta_\delta}\left[\mathcal{Z_\delta}^2(X)-A^2x_f^{4-2
\delta-4\eta_\delta}e^{-2x_f}\right].
\label{e46}
\end{equation}
The criterion of dominant balance requires that
\begin{equation}
\frac{1}{\kappa}=\lambda x_f^{-n-\frac{5}{2}+\delta+2\eta_\delta}.
\label{e47}
\end{equation}
Following earlier arguments [see (\ref{e13})], in the inner region we have
$$\frac{d}{dX}\mathcal{Z_\delta}(X)\approx-\mathcal{Z_\delta}^2(X).$$
The solution to this equation is
\begin{equation}
\mathcal{Z}_\delta(X)=1/\left(X+\mathcal{D}_\delta\right),
\label{e48}
\end{equation}
where $\mathcal{D}_\delta$ is a constant of integration. Matching (\ref{e44})
with (\ref{e48}) gives
\begin{equation}
\mathcal{D}_\delta=\frac{1}{\mathcal{C}_\delta}-\frac{\lambda}{\left(n+3/2-
\delta-2\eta_\delta\right)x_f^{n+3/2-\delta-2\eta_\delta}}.
\label{e49}
\end{equation}

As in (\ref{e19}), the matching of the solutions in the equilibrium and
freeze-out regions determines that
\begin{equation}
\mathcal{D}_\delta=\lambda x_f^{-n-5/2+2\eta_\delta+\delta}.
\label{e50}
\end{equation}
The analog of (\ref{e22}) is
\begin{equation}
\frac{1}{\mathcal{C}_\delta}=\lambda\frac{n+3/2-2\eta_\delta-\delta+x_f}{n+3/2-
2\eta_\delta-\delta}\,x_f^{-n-5/2+2\eta_\delta+\delta}.
\label{e51}
\end{equation}
Here, $x_f=x_{f,ns}$. Consequently, for $\phi=1/2-\delta$ and $\delta$ small,
the large-$x$ asymptotic behavior of $Y(x)$ is
\begin{equation}
Y(x)\sim C_\delta x^{-1/2+2\zeta_\delta+\delta}.
\label{e52}
\end{equation}
In such a case $C_\delta$ and the freeze-out region are determined from
(\ref{e51}) and from (\ref{e10}), while the freeze-out point is given by
(\ref{e11}). We note that the limit $\delta\to0$ is smooth.

\subsection{The case of general $\phi>0$ with $\phi$ not near $1/2$}

By the arguments given in Sec.~\ref{sec:2} for large $x$ in the post-freeze-out
region the approximate solution to (\ref{e36}) is $Z(x,\phi)\approx Z^{\rm
post-freeze-out}(x,\phi)$, where
\begin{equation}
\frac{d}{dx}Z^{\rm post-freeze-out}(x,\phi)=-\lambda f(x,\phi)\left[Z^{\rm
post-freeze-out}(x,\phi)\right]^2.
\label{e53}
\end{equation}
The solution to this equation is
\begin{equation}
Z^{\rm post-freeze-out}(x,\phi)=\frac{1}{C_\phi^{-1}+\lambda\int dx\,f(x,\phi)},
\label{e54}
\end{equation}
where $C_\phi$ is a positive constant. Equation (\ref{e54}) is valid for general
$\phi>0$.

It is convenient to rewrite (\ref{e34}) and (\ref{e37}) using the function
\begin{equation}
h(x,\phi)\equiv{(1-x^{1-2\phi})/(1-2\phi)}.
\label{e55}
\end{equation}
We then have
$$g(x,\phi)=x^\phi\exp\left(\frac{9\zeta_{\phi}}{2\phi}x^{-2\phi}\right)\exp
\left[2\zeta_\phi h(x,\phi)\right]$$
and
$$f(x,\phi)=x^{-n-2-\phi}\exp\left(-\frac{9\zeta_{\phi}}{2\phi}x^{-2\phi}\right)
\exp\left[-2\zeta_{\phi}h(x,\phi)\right].$$
In the limit as $x\to0$
$$h(x,\phi)\to\left\{\begin{array}{cl}\frac{1}{1-2\phi},&\mbox{for
\ensuremath{\phi<1/2}},\\
-\frac{1}{2\phi-1}, & \mbox{for \ensuremath{\phi>1/2}},
\end{array}\right.$$
and so $f(x,\phi)\to0$. Furthermore as $x\to\infty$, $Z_{\rm eq}(x,\phi)$ is
negligible because in this limit
$$h(x,\phi)\to\left\{\begin{array}{cl} -\infty & \mbox{for \ensuremath{\phi<
1/2}},\\ -\frac{1}{2\phi-1} & \mbox{for \ensuremath{\phi>1/2}}.
\end{array}\right.$$

\subsubsection{The case $0<\zeta_\phi<\phi\ll 1/2$}

Next, we consider the case for which $\zeta_\phi/\phi\sim O(1)$ and $\zeta_\phi
x_f\ll 1$. This case illustrates the competition between space-time foam and
dilaton sources in their effect on the relic abundance. In this case
\begin{eqnarray}
g(x,\phi) &\sim& x^\phi\exp\left(\frac{9\zeta_\phi}{2\phi}\right),\nonumber\\
f(x,\phi) &\sim& x^{-n-2-\phi}\exp\left(\frac{9\zeta_\phi}{2\phi}\right),
\label{e56}
\end{eqnarray}
for $x$ in the freeze-out region and $x\gg x_f$. The analog of (\ref{e5}) is
similar except that $\lambda$ is replaced by $\lambda\exp\left(-\frac{9
\zeta_\phi}{2\phi}\right)$.

The analog of (\ref{e10}) is
\begin{equation}
x_f\sim\log\left[2A\lambda\exp\left(-\frac{9\zeta_\phi}{2\phi}\right)\right]-
(n+1/2)\log\left(x_f\right)=\log(2A\lambda)-(n+1/2)\log\left(x_f\right)-\frac{9
\zeta_\phi}{2\phi}.
\label{e57}
\end{equation}
The previous analysis then implies that for large $x$ we have the following
asymptotic behavior for $Y(x)$:
\begin{equation}
\label{e58}
Y(x)\sim\frac{(n+1+\phi)x_f^{n+2+\phi}}{\lambda(n+1+\phi+x_f)}x^{-\phi},
\end{equation}
where
\begin{equation}
\label{e59}
x_f=(n+1/2)W\left(\left[2\lambda A\exp\left(-\frac{9\zeta_{\phi}}{2\phi}\right)
\right]^{n+1/2}\Big /(n+1/2)\right).
\end{equation}
We denote this value of $x_f$ by $x_{f,1}$. The scaling (\ref{e58}) is formally
similar to the pure time-dependent dilaton case in (\ref{e23}), but the effects
of the D-foam are incorporated only in the shifted value of the freeze-out point
$x_f$ in (\ref{e59}).

\subsubsection{The case $\phi\gg1/2$}

For $\phi\gg 1/2$ and $x\gg x_f$ we have
\begin{equation}
\frac{1}{Z}=-\lambda\exp\left(\frac{2\zeta_\phi}{2\phi-1}\right)\frac{x^{-
n-1-\phi}}{n+1+\phi}+\frac{1}{\mathcal{C}},
\label{e60}
\end{equation}
where $\mathcal{C}$ is a constant. To leading order the analog of (\ref{e9})
for this case is independent of $\phi$ and $\zeta_\phi$, so $x_f$ is determined
by (\ref{e10}). In the inner (boundary-layer) region we again write $x=x_f+
\kappa X$ and $\mathcal{Z\left(\mathit{X}\right)=\mathit{Z\left(x_f+\kappa X
\right)}}$. Hence,
$$\frac{1}{\kappa}\frac{d\mathcal{Z}}{dX}\simeq-\lambda\left(x_f+\kappa X
\right)^{-n-2-\phi}\exp\left(\frac{2\zeta_\phi}{2\phi-1}\right)\left[\mathcal{
Z}^2(X)-Z_{\rm eq}^2\left(x_f\right)\right],$$
where $Z_{\rm eq}^2\left(x_f\right)=\frac{1}{4\lambda^2}x_f^{4+2\phi+2n}\exp
\left(-\frac{4\zeta_\phi}{2\phi-1}\right)$. The principle of dominant balance
then implies that
$$\frac{1}{\kappa}=\lambda x_f^{-n-2-\phi}\exp\left(\frac{2\zeta_\phi}{2\phi-1}
\right).$$
Hence, $\frac{d\mathcal{Z}}{dX}=-\mathcal{Z}^2$ with the solution $\mathcal{Z}
(X)=1/(x+\mathcal{D})$, where $D$ is a constant.

Matching the equilibrium region to the boundary layer gives
$$2Z_{\rm eq}\left(x_f+\kappa X\right)\approx 1/(x+\mathcal{D}).$$
This implies that
\begin{equation}
\mathcal{D}=\lambda x_f^{-n-2-\phi}\exp\left(\frac{2\zeta_\phi}
{2\phi-1}\right).
\label{e61}
\end{equation}
Matching the freeze-out-region solution to the post-freeze-out-region
solution (\ref{e60}), we find that
\begin{equation}
\frac{1}{\mathcal{C}}=\lambda\exp\left(\frac{2\zeta_\phi}{2\phi-1}\right)
x_f^{-n-2-\phi}\left(1+\frac{x_f}{n+1+\phi}\right).
\label{e62}
\end{equation}
Finally, we find that as $x\to\infty$,
\begin{equation}
Y(x)\sim\frac{1}{\lambda}\,\exp\Big(-\frac{2\zeta_\phi}{2\phi-1}\Big)\frac{x^{-
\phi}}{x_f^{-n-2-\phi}+\left(x_f^{-n-1-\phi}-x^{-n-1-\phi}\right)/(n+1+\phi)}.
\label{e63}
\end{equation}
and $x_f=x_{f,ns}$.

\subsubsection{The approach to $\phi=0$}

The integrating factor in (\ref{e34}) is singular as $\phi\to 0^+$. However, the
function $g$ is only determined up to an $x$-independent factor. In order to
study the limit $\phi\to0$ we consider a modified $g(x,\phi)$ and an associated
$f(x,\phi)$, which we denote $\widetilde{g}(x,\phi)$ and $\widetilde{f}(x,\phi)$
respectively. These functions have the following form:
\begin{equation}
\widetilde{g}(x,\phi)=x^\phi\,\exp\left[2\zeta_\phi h_1(x,\phi)\right]\,
\exp\left[9\zeta_\phi h_2(x,\phi)\right],
\label{e64}
\end{equation}
where
\begin{equation}
h_1(x,\phi)\equiv\frac{1-x^{1-2\phi}}{1-2\phi},\qquad
h_2(x,\phi)\equiv\frac{x^{-2\phi}-1}{2\phi}-\frac{2}{9},
\label{e65}
\end{equation}
and, as before, we have the relation
$$\widetilde{f}(x,\phi)\equiv x^{-n-2}/\widetilde{g}(x,\phi).$$

The limits $x\to\infty$ and $\phi\to0$ do {\it not} commute (a feature that is
common to other limits involving $x$). Parallel to the discussion of
Ref.~\cite{R8}, we take the limit $\phi\to0$ first. It is straightforward to
show that for large $x$ but $\zeta x$ still small one obtains
\begin{equation}
1/\mathcal{C}=\mathcal{D}+\lambda x_f^{-n-1+9\zeta}/(n+1-9\zeta).
\label{e66}
\end{equation}
By matching the equilibrium region to the freeze-out region we obtain
\begin{equation}
\mathcal{D}=\lambda x_f^{-n-2+9\zeta}.
\label{e67}
\end{equation}

These formulas are similar to the the case of dilaton cosmology in the absence
of space-time foam with the crucial difference that $\phi$ is now
replaced by $-9\zeta$. Finally, we obtain
\begin{equation}
Y(x)\sim\frac{x^{9\zeta}}{\mathcal{C}^{-1}-\lambda x^{-n-1+9\zeta}/(n+1-9\zeta)}
,
\label{e68}
\end{equation}
which indicates the role of D foam as a source of particle production in this
case, in the sense that $Y$ increases as $x$ increases. Also, in this case $x_f=
x_{f,ns}$. Notice that the behavior (\ref{e67}), which indicates an increase of
the DM thermal relic abundance with decreasing temperature, is compatible with
our earlier numerical investigations in \cite{R8}. For $x\gg x_f$ (as in the
current universe) the abundance (\ref{e68}) can be approximated by
\begin{equation}
Y(x)\sim\lambda^{-1}x_f^{n+2}\left(x/x_f\right)^{9\zeta}\quad(x\gg x_f),
\label{e69}
\end{equation}
which we use in the Sec.~\ref{sec:4} to discuss the phenomenology of these
models.

\section{Phenomenological implications}
\label{sec:4}

As mentioned earlier, the phenomenologically relevant quantity that can be
compared directly with experiments is the Hubble-constant-free relic abundance,
$\Omega h^2=m\mathcal{N}/\rho_0^c$, where $\rho_0^c$ is the current critical
density and $\mathcal{N}$ is the number density of the DM species. For DM
species $X$ with mass $m_X$ this quantity is given by (\ref{e38}) \cite{R3}. The
behavior of $\Omega_X$ is then readily obtained for all cases studied in this
work.

The analysis in the previous sections indicates that time-dependent sources in
our cosmological models lead to modified relic abundances for DM species, as
compared to those computed within the standard cosmology. This modification can
be quantified by considering the
ratio (\ref{e39}) in which the numerator and denominator may involve different
freeze-out temperatures. Since both expressions are known theoretically, the
ratio (\ref{e39}) is computable explicitly for all cases studied above.

Before proceeding with the phenomenology of the various sources discussed in
this article, we make some generic remarks. If the sources are such that there
is \emph{dilution} of DM relic abundance \emph{relative} to the prediction of
standard cosmology, this can have important phenomenological implications for
new physics, such as supersymmetry (SUSY) at colliders~\cite{R5,R6,R7}. In such
a case there is a \emph{larger} portion of the available parameter space of the
SUSY model, which is compatible with the WMAP and other
cosmological/astrophysical data~\cite{wmap}.

More room for supersymmetry implies heavier partners, which in turn may have
interesting signatures at colliders, such as the Large Hadron Collider (LHC).
If the relic density of the neutralino ${\tilde \chi}^0_1$, which is the dominant DM in
SUGRA-like models, is diluted by a factor of about $1/10$ in the presence of sources, then the
final states expected at the LHC consist of Z-bosons, Higgs bosons and  $\tau$-leptons. Such
states are produced when one looks at the decay chains
of the dominant SUSY production mechanism of squark $\tilde q$ and gluino
$\tilde g$ pairs at the LHC:
$$\tilde q \,\to\,q\,\tilde\chi_2^0\,\to\,q\tau\,\tilde\tau_1\,\to\,q\tau\tau{\tilde
\chi}^0_1,\quad{\tilde\chi}^0_2\,\to\,h^0\,{\tilde\chi}^0_1,\quad{\tilde
\chi}^0_2\,\to\,Z\,{\tilde\chi}^0_1,$$
where ${\tilde\chi}^0_2$ is the next-to-lightest neutralino, and $h^0$ is the
Higgs particle. In Ref.~\cite{R7} a detailed analysis in the standard parameter
space $m_{1/2},m_0$ (where $m_{1/2}$ and $m_0$ are the gaugino and scalar
masses) of mSUGRA models has been performed. In this analysis the parametric
regions for the dominant decay patterns at the LHC:
\begin{enumerate}
\item Higgs+jets+missing~transverse~energy,
\item Z+jets+missing~transverse~energy,
\item 2$\tau$+jets+missing~transverse~energy,
\nonumber
\end{enumerate}
have been predicted.
Dilution factors of about 1/100 or more are compatible with the analysis in this
paper for reasonable values of the parameters. Such dilutions may even push the
parameter spaces of minimal supersymmetric models beyond the reach of the LHC
(assuming standard-model-like Higgs particle masses of about 125 GeV). For
instance, in the constrained minimal supersymmetric standard model (CMSSM) with Higgs-mass range
123 - 128 GeV and $\tan \beta$ of about 50, Lahanas and Spanos discussed the dilaton-induced
dilution factor \cite{R6}. On including the effect of the dilution factor, they showed that the constraints placed on the
parameter space of CMSSM, from the current ATLAS and CMS SUSY searches for DM, were not sufficient to exclude the model.

We proceed to discuss the phenomenology of the cases discussed above by giving
the corresponding values of the ratio (\ref{e39}) today. We assume that the
freeze-out points $x_{f,ns}$ in the absence of sources are about 30 , as
expected in typical phenomenological models in which the DM is identified as a
supersymmetric partner, such as a neutralino.

The temperature of the universe, which is used in the definition of $x$ today
$x_0$, is that of the cosmic microwave background (CMB) temperature $T_{\rm
CMB}\approx2.35\times 10^{-13}$~GeV. Thus, for DM masses in the range $m_X
\approx$ 300 GeV - 1~TeV, we have
\begin{equation}
x_0\equiv m_X/T_{\rm CMB}\approx 10^{15}-10^{16}.
\label{e70}
\end{equation}
Moreover, we assume that the source-free relic abundance $Y_{ns}(x)$ currently,
which approaches a constant as $x\to\infty$ \cite{R3}, as the boundary-layer
analysis of Ref.~\cite{R1} confirms, is given by
\begin{equation}
\lim_{x\to\infty}Y_{ns}\approx\frac{(n+1)\,{x^{n+2}_{f,ns}}}{\lambda\,(n+1+
x_{f,ns})}.
\label{e71}
\end{equation}

Recall that the freeze-out point in the source-free case $x_{f,ns}$ indicates a
range of values of $x$ in the vicinity of (\ref{e10}) with $A\approx0.000145$
\cite{R1}. For all but the case $0<\zeta_\phi<\phi\ll 1/2$ the freeze-out point
$x_f=x_{f,ns}$. However, as is evident from (\ref{e57}), even in the case $0<
\zeta_\phi<\phi\ll 1/2$, the freeze-out point is shifted by an amount less than
9/2: $x_f^{\phi\ll 1/2}\sim x_{f,ns}-\frac{9\zeta_\phi}{2\phi}$. In the models
we consider here $x_{f,ns}\approx30$, so such a shift is not significant. Thus,
from now on we treat $x_f\approx x_{f,ns}$ in all cases. This simplifies the arguments
and allows an easy estimate of the ratio $\mathcal{R}$ in (\ref{e39}).

As a starting point, we take the case of a time-dependent dilaton source of the
form (\ref{e3}) in the absence of D-foam; that is, $\zeta_\phi=0$. This case was
discussed in Refs.~\cite{R5,R6} and was revisited in Ref.~\cite{R1} using
asymptotic matching techniques. From (\ref{e23}) and (\ref{e71}) the ratio
(\ref{e39}) becomes (upon setting $x_f\sim x_{f,ns}$)
\begin{equation}
\mathcal{R}^{\rm dilaton}(x=x_0)\sim\frac{n+1+\phi}{n+1}\,\frac{n+1+x_{f,sn}}
{n+1+\phi+x_{f,ns}}\left(x_{f,ns}/x_0\right)^\phi
\label{e72}
\end{equation}
with $x_0$ given in (\ref{e70}).

From (\ref{e58}) we then notice that (\ref{e72}) also applies to
the case of nontrivial D-foam but with $0<\zeta_\phi<\phi\ll 1/2$. For $x_{f,sn}$
about $30$ and for $s$-wave scattering ($n=0$) the
approximate thermal DM relic dilution factor (\ref{e72}) is determined by
$\left(x_{f,ns}/x_0\right)^\phi\approx 10^{-16\phi}$ for DM masses $m_x$ in the range 0.3 , 1~TeV.
 Thus, to obtain a dilution factor of order $1/10$, which is relevant for LHC
phenomenology, we need values of $\phi$ near $1/16$,
which is small compared with $1/2$ and which is consistent. However, the case of
phenomenologically significant dilution requires that $\zeta_\phi x_f\ll 1$ and
thus $\zeta_\phi\ll\phi$. For the pure dilation case, in the absence of D foam,
one may have larger values of $\phi$ that lead to acceptable phenomenology; for
instance, a dilution of about $10^{-2}$ can be obtained with $\phi
\approx1/8$.

On the other hand, in the case where the space-time defect (D-foam defect)
dominates the time-dependent dilaton effect, that is when the strength of the
foam fluctuations is such that $\zeta\gg\phi\to0$, we have an \emph{enhancement}
of the DM relic abundances rather than a dilution as the temperature
decreases. This becomes clear from (\ref{e68}) and (\ref{e69}). In such cases
there is \emph{less room} for supersymmetry available in the relevant parameter
space as compared with the source-free case after cosmological (WMAP) constraints~\cite{wmap} are taken into account.

The enhancement factor scales like
\begin{equation}
\mathcal{R}(x=x_0)\sim\frac{n+1+x_f}{n+1}\left(x_0/x_f\right)^{9
\zeta}.
\label{e73}
\end{equation}
For $s$-wave scattering and with $x_0$ given by (\ref{e70}) this implies that $\mathcal{R}\sim\big(10\big)^{(136\,-\,154)\zeta}$. Such
models lead to more severe constraints on the available supersymmetry parameter
space if the enhancement is observable.

Therefore, for these models to be phenomenologically viable today, this requires
$\mathcal{R}$ to be $\mathcal{O}(1)$ within experimental error, so that the increase
compared to the source-free (standard) case is not appreciable. This requires
that $\zeta<10^{-3}$, so that the error in calculating abundances would match
the per mil level of the current errors in experimental
astrophysical measurements \cite{wmap}. Because of (\ref{e33}), this implies
that
\begin{equation}
\zeta=2x_0\left(g_0m_X/M_s\right)^2\langle\Delta^2\rangle_0<10^{-3}.
\label{e74}
\end{equation}
For the range of $x_0$ in (\ref{e70}) this is satisfied for heavy D particles of
masses $M_s/g_0\ge\left(10^{11}\,-\,10^{13}\right)/\langle\Delta^2
\rangle_0\,{\rm GeV}$, and $m_X$ in the range $0.3$ - 1 TeV and $\langle
\Delta^2\rangle_0\le1$.

Next, we discuss the case of a time-dependent dilaton with $\phi\gg 1/2$ in
(\ref{e63}). Since we always assume weak foam fluctuations, $\zeta_\phi<1$, this
case is also dilaton dominated, and hence we expect a significant dilution of
the relic abundance. Indeed, because of (\ref{e63}) and (\ref{e71}), in
this case we obtain the following expression in the limit $x\to x_0\gg x_f=
x_{f,ns} $ for the ratio $\mathcal{R}$ in Eq.~(\ref{e39}):
\begin{equation}
\mathcal{R}\sim\exp\Big(-\frac{2\zeta_\phi}{2\phi-1}\Big)\Big(\frac{n+1+\phi}
{n+1}\Big)\Big(\frac{n+1+x_{f,ns}}{x_{f,ns}}\Big)\,\left(x_f/x_{f,ns}\right)^{n+
1}\left(x_f/x\right)^\phi,
\label{e75}
\end{equation}
where we assume that the freeze-out points $x_f\sim x_{f,ns}$ are about $30$ or
more. Thus, we have significant dilution of the DM relic densities at late
epochs of the universe. For instance, in the present era and for DM masses in the
range $m_X \approx 300$~GeV, the ratio is $\mathcal{R}\approx\exp\left(-\frac{2
\zeta_\phi}{2\phi-1}\right)\times(\approx 200)\times 10^{-15\phi}$ for
s- or p-wave scattering ($n=1,2$). Thus, we see that for $\zeta \ll 1$, which is
natural in the case of D foam with D particles whose masses are higher than a TeV
\cite{R8}, the main factor that drives the dilution is the value of the dilaton
parameter $\phi$.

In the case $\phi\approx 5$, for instance, the dilution factor is already enormous
(it is of order $10^{-75}$), so in such models practically all DM today will
have disappeared. This may rule these models out phenomenologically, although the
situation with DM and its nature is currently unclear, as there is no concrete
evidence for it apart from the galactic motion. For this reason alternative
theories with no dark matter but modified gravity at galactic scales have been
considered extensively in the literature. We do not consider them here,
since in our opinion the evidence against them,
especially from galactic lensing measurements, is significant. Thus, all we can
say is that this type of supersymmetric DM (satisfying
$x_f\approx 30$) would be diluted in this model, and it would be practically
absent today. Other types of DM that would not couple to the dilaton, might
survive.

Next we discuss the cases for which $\phi$ is near $1/2$. Now, using
(\ref{e51}) and (\ref{e52}), we obtain for $x\to x_0\gg x_f\sim x_{f,ns}$:
\begin{equation}
\mathcal{R}\sim\frac{n+3/2-2\zeta_{1/2-\delta}-\delta}{n+3/2-2\zeta_{1/2-\delta}
+x_{f,ns}}\,\frac{n+1+x_{f,ns}}{n+1}\,\left(x_{f,ns}/x\right)^{1/2-2\zeta_{1/2-
\delta}-\delta}.
\label{e76}
\end{equation}
We observe that for $\delta\to0$ and $\zeta_{1/2}<\phi=1/2$, the main dilution
comes from the dilaton effects and scales with $x$ as $(x_{f,ns}/x)^{1/2-
\zeta_{1/2}}$. Thus the dilution due to the dilaton is compensated
by foam fluctuation effects, so for $\zeta_{1/2}=\phi/2=1/4<\phi=1/2$
there is no appreciable dilaton-driven dilution, and the ratio (\ref{e76}) tends
to one ($R_{\zeta_{1/2}=1/4}\approx1$) for any $n>0$.

We observe from (\ref{e33}) that $\zeta_{1/2}$ is independent of $x_0$:
\begin{equation}
\zeta_{1/2}=2g_0^2\langle\Delta^2\rangle_0m_X^2/M_s^2
\label{e77}
\end{equation}
and the condition $\zeta=1/4$ implies that $\langle\Delta^2\rangle_0=M_s^2/
\left(8g_0^2\,m_X^2\right)<1$, where the inequality on the right side ensures
naturalness in the fluctuations of a weak foam, which we have assumed
throughout. The latter condition necessitates $m_X\approx M_s/g_s$. We stress
that in this case the result for the relic abundance today turns out to be
equal to the standard source-free case independent of the actual freeze-out
point, and hence in principle $m_X$ is only constrained to be of the same order
of magnitude as the D-particle mass $M_s/g_0$.

Finally, we mention that one may consider a $\delta\ge1/8$ to produce dilution
of order $\mathcal{R}\le{10^{-2}}$ in
the relic abundance (\ref{e76}), thereby opening the possibility of pushing
this class of supersymmetric models out of the reach of the LHC, according to
the analysis in Ref.~\cite{R6}. However, in this case, (\ref{e33}) implies that
the condition $1/4=\zeta_\delta\sim2g_0^2\,x_0^{-\delta}\langle\Delta^2\rangle_0
m_X^2/M_s^2$ for $x_0$ in the region (\ref{e70}) can be satisfied for $g_0^2
m^2_X/M^2_s\sim(1/8)10^{15\delta}\langle\Delta^2\rangle_0$. To ensure that $m_X
\le M_s/g_0$ this would imply naturally small fluctuations in the foam $\langle
\Delta^2\rangle_0\sim 10^{-15\delta}$ with $\delta\ge1/8$.

The above predictions are quite generic and hence they are largely independent
of the details of the underlying microscopic model. Nevertheless, the cosmology
of the models, in particular the precise dependence of the dilaton on the cosmic
time at various eras of the universe, is an open issue. The lack of detailed
microscopic models that would determine the form of the dilaton potential and
provide rigorous information on the region of validity of the dilaton
cosmological solution (\ref{e3}) complicates matters. Nevertheless, one
may perform phenomenological searches on the compatibility of such solutions at
various epochs of the universe. For the DM searches mentioned above, all one
needs is the dominance of the time-dependent dilaton at early epochs of the
universe before the big-bang nucleosynthesis. Nevertheless, a dilaton of
the form (\ref{e3}) can be compatible (notably at the same level as the
$\Lambda$CDM model) with cosmological data even at low redshifts of order $z=O
(1)$, where large scale structure in the universe (galaxies and clusters of
them) is formed, as demonstrated recently in Ref.~\cite{R12}. On the other hand,
D-foam dominance at late eras (such as the end of radiation or matter-dominated era
\cite{yusaf}) has been argued to play a role in galactic growth itself.
Thus, considering models with combined dilaton and D-foam sources, as in the
current article, may be desirable from the point of view of constructing
realistic cosmologies in such frameworks. However, the rate of galactic growth
is in principle capable of discriminating the various models (\ref{e3})
corresponding to different values of $\phi$ when more data become available in
the near future. In all such theories, of course, an important requirement is
that the big-bang-nucleosynthesis conditions at MeV temperatures are not disturbed.

\acknowledgments
CMB thanks the Leverhulme Trust (U.K.) and the U.S.~Department of Energy for
financial support. The work of N.E.M. is supported in part by the London Centre
for Terauniverse Studies (LCTS), using funding from the European Research
Council via the Advanced Investigator Grant 267352. The work of NEM and SS is
also supported in part by the U.K.~Science and Technology Facilities Council
(STFC).

\section*{APPENDIX: Thermodynamic properties of a universe in the presence of
sources}

The purpose of this appendix is to demonstrate that it is possible to define an appropriate entropy density (scaling with temperature as $T^3$),  even in the presence of nontrivial backgrounds, such as a time-dependent dilaton
and/or space-time foam.  This allows the entropy density to be used in this paper as a fiducial quantity in the definition of the thermal relic abundance $Y(x)$.

In the presence of such nontrivial backgrounds the continuity equations of cosmic fluids
corresponding to matter and radiation are modified relative to standard
Friedmann-Robertson-Walker (FRW) cosmology. These modifications could affect the
thermodynamic properties of the universe, such as the relation between the scale
factor and the temperature $T$ (the cooling law). It is the \emph{relativistic} degrees of freedom that
dominate the entropy and the cooling law. In the case of a FRW universe, the
continuity equation is the conservation of the stress-energy tensor $\nabla^\mu
T_{\mu\nu}=0$, which is compatible with Einstein's equations of general
relativity and admits a thermodynamic interpretation. This equation can
be manipulated to appear as the first law of
thermodynamics for the total internal energy $\rho V$ in a co-moving volume $V\sim
a^3$ and pressure $p$:
\begin{equation}
d(\rho V) + pdV=0,\quad V\sim a^3.
\label{e78}
\end{equation}

This interpretation in terms of the first law is consistent with an \emph{adiabatic} expansion at temperature $T$. A constant entropy function $S(T,V)$, analytic in   $T$ and  $V$, can be constructed:
\begin{equation}
S=V\frac{\rho+p}{T}.
\label{e79}
\end{equation}
The construction involves the application of the thermodynamic Maxwell relations
to cast the right side of (\ref{e78}) into the form $TdS$; that is\footnote{Strictly speaking Eq.~(\ref{e80}) should be modified by a term involving
the chemical potential $\mu$; hence $T dS$ should be replaced by  $TdS + \mu dN$. As is standard in cosmology $\mu $ is ignored   because $\mu/T$ is much smaller than one~\cite{R3}, which is consistent with the dominance of the relativistic degrees of freedom in the entropy.},
\begin{equation}
0=TdS=d(\rho\,V) + pdV.
\label{e80}
\end{equation}
The dominance of relativistic degrees of freedom in the entropy $S$ is therefore
consistent with the constancy of $S$ and the cooling law $a\sim 1/T$. (Recall
that $\rho=3p\sim T^4$ for radiation.) From (\ref{e79}) and (\ref{e80}) it is
then straightforward to see that the entropy density $s\equiv S/V$ scales with
temperature $T$ as $T^3$.

For cosmologies with nontrivial time-dependent dilaton and/or space-time
D-particle foam backgrounds we also construct entropy functions that are constant during the evolution of the universe.

\bigskip
{\bf (i) Dilaton Cosmology}
\medskip

For a FRW cosmology in four space-time dimensions the presence of a rolling
dilaton leads to a modification of the continuity equation for the energy
density $\rho$ and pressure $p$ \cite{R4,R5,R6}:
\begin{equation}
{\dot\rho}+3H(\rho+p)-{\dot\Phi}(\rho-3p)=0,
\label{e81}
\end{equation}
where the dot denotes derivative with respect to the cosmic time $t$.
(Here $\rho$ and $p$ here denote the total energy density of the fluid.)
We note that:
\begin{itemize}
\item
From (\ref{e81}) the dilaton source terms do not
play a role for radiation; one obtains the standard scaling of $\rho\sim a^{-4}$ in the
radiation dominated era of the universe.
\item
For dust, $p=0$. Also, for DM with mass $m_X$,
$\rho_X=m_X n_X$, where $n_X$ is the number density; the source-independent part
of (\ref{e81}) yields the collisionless Boltzmann equation for thermal relic
abundance. The dilaton-dependent term is a classical source term ${\dot\Phi}n$.
\end{itemize}

The nonlinear part of the Boltzmann equation comes from two-body annihilations
of DM particles. On assuming the functional dependence $\rho=\rho(a)$ and a
dilaton source of the form (\ref{e3}), we obtain from (\ref{e81})
\begin{equation}
d(\rho V)+pdV-|\Phi_0|(\rho-3p)dV/3=0.
\label{e82}
\end{equation}
Here we have taken into account that ${\dot
\Phi}=-|\Phi_0|H$, where $H={\dot a}/a$ is the Hubble parameter,
$V\sim a^3$ is the co-moving volume, and $\rho a^3$ is the total
(internal) energy in that volume.

We thus
observe that the presence of a rolling dilaton affects the standard
thermodynamic properties of the FRW universe. Our aim is to ascertain whether
the total entropy in the co-moving volume $V$ remains constant in time after the
inclusion of the dilaton source (\ref{e3}).
A naive application of the first law of thermodynamics would identify $d(\rho V)
+pdV$ with $TdS$, where $T$ is the temperature, and $S$ is the total entropy in
the volume $V$. It would seem that a dilaton source leads to the nonconservation
of entropy. However, this is incorrect. To show this, we first replace the zero
of the right side of (\ref{e82}) by $Td\mathcal{S}$, where $\mathcal{S}$ is the
quantity that represents the entropy; the entropy is assumed to
depend on $T$ and $V$, so $\mathcal{S}=\mathcal{S}(T,V)$. From (\ref{e82}) we
find that
\begin{equation}
d\mathcal{S}=\frac{1}{T}d(\rho V)+\frac{p}{T}dV-\frac{|\Phi_0|}{3T}(\rho-3p)dV=
V\frac{d\rho}{dT}dT+\frac{1}{T}\left[(1+|\Phi_0|)p+(1-|\Phi_0|/3)\rho\right]dV.
\label{e83}
\end{equation}
As in the case of conventional cosmology, it has been assumed that the cooling law of (\ref{e27}) holds, and that $\rho=\rho(T)$ and $p=p(T)$, since both $\rho$ and $p$
depend upon the scale factor $a$, which is a function of temperature, $a=a(T)$.
The function $\mathcal{S}(T,V)$ is assumed to be a differentiable function of
$T$ and $V$. This implies the condition:
\begin{equation}
\frac{\partial^2\mathcal{S}}{\partial T\partial V}=\frac{\partial^2\mathcal{S}}
{\partial V\partial T}.
\label{e84}
\end{equation}
From (\ref{e83}) and (\ref{e84}) we then obtain
\begin{equation}
\frac{1}{T^2}\big[(1+|\Phi_0|)p+(1-|\Phi_0|)/3)\rho\big]=\frac{(1+|\Phi_0|)}{T}
\frac{dp}{dT}-\frac{|\Phi_0|}{3T}\frac{d\rho}{dT}.
\label{e85}
\end{equation}

The expression (\ref{e83}) for $d\mathcal{S}$ can be rewritten as
\begin{eqnarray}
d\mathcal{S} &=& \frac{1}{T}d\left[(\rho+p)V\right]-\frac{V}{T}\frac{dp}{dT}
dT-\frac{|\Phi_0|}{3\,T}\,d\left[(\rho-3p)V\right]+\frac{|\Phi_0|V}{3T}\,\frac{d
\rho}{dT}\,dT-\frac{|\Phi_0|V}{T}\,\frac{dp}{dT}\,dT\nonumber\\
&=& d\left[\frac{1}{T}\left(\rho\left[1-\frac{|\Phi_0}{3}\right]+p\left[1+|\Phi_0|\right]\right)V
\right]+\frac{V}{T^2}\left[(1+|\Phi_0|)p+(1-\frac{|\Phi_)}{3})\rho\right]dT-
\frac{(1+|\Phi_0|)V}{T}\frac{dp}{dT}dT+\frac{|\Phi_0|V}{3T}\frac{d\rho}{dT}dT
\nonumber\\
&=& d\left[\frac{1}{T}\left(\rho\left[1-\frac{|\Phi_0}{3}\right]+p\left[1+|\Phi_0|\right]\right)V\right].
\label{e86}
\end{eqnarray}
In the last equality on the right side we have used (\ref{e85}). From
(\ref{e86}) we conclude that the quantity
\begin{equation}
\mathcal{S}(T,V)\equiv[\rho(1-|\Phi_0|/3)+p(1+|\Phi_0|)
]V/T
\label{e87}
\end{equation}
is constant upon using the classical equations of motion [or equivalently,
the continuity equation (\ref{e81}) for the case of dilaton
cosmology (\ref{e3})]. $\mathcal{S}$ may be identified with the total entropy in the
co-moving volume $V$. The corresponding entropy density $s$ is then:
\begin{equation}
s=\frac{1}{T}\big[\rho(1-|\Phi_0|/3)+p(1+|\Phi_0|)\big],
\label{e88}
\end{equation}
which, in view of (\ref{e87}), scales with the size of the universe as
$a^{-3}=(T/C_0)^3$, upon assuming (\ref{e27}).
 We stress
that the energy density $\rho$ and pressure $p$ in the above formulas pertain to the \emph{total} degrees of freedom of the fluid, including the relativistic ones.
It is the latter, for which the dilaton source effects are \emph{irrelevant} [see (\ref{e81})], that provide the dominant contributions to the entropy; otherwise, the entropy would not remain constant. Indeed, in the case of DM dust, $p=0$,
the entropy density is $s=\rho(1-|\Phi_0|/3)/T$, which does not leave the entropy function (\ref{e87}) constant. This is satisfied only for relativistic degrees of freedom that have an energy density scaling like $\rho \sim T^4$ with the temperature $T$.

We have the following relation between $Y$ and the number density $n_X$ of the DM species X:
\begin{equation}
Y=n_XT^{-3}\to n_X=m_X^3 Yx^{-3},
\label{e89}
\end{equation}
as in the standard cosmology case. The energy density $\rho_X$ of the DM relic
satisfies $\rho_X=m_Xn_X$. The current relic abundance:
\begin{equation}
\Omega_Xh^2\sim\frac{m_X^4}{\rho_0^c}\frac{Y_0}{x_0^3},
\label{e90}
\end{equation}
occurs for $x=x_0=m_X/T_0$, with $T_0$ the current (CMB) temperature of the universe
and $\rho_0^c$  the current critical density. Since the latter is
proportional to $h^2$, the above expression is independent of the value of the
Hubble-constant.

In practice, $x_0\gg 1$. Hence, the asymptotic regime $Y(x)$ with $x\to\infty$ is
relevant. In the current literature one usually replaces $Y_0$ by $Y_\infty$;
that is,
\begin{equation}
\Omega_X h^2\sim\frac{m_X^4}{\rho_0^c}\frac{Y_\infty}{x_0^3}.
\label{e91}
\end{equation}
For standard cosmology in (\ref{e62}) $\lim_{x\to\infty}Y(x)={\rm
constant}$. This constant value of the freeze-out  is identified with the current relic
abundance of the weakly-interacting massive particle (WIMP) $\Omega_Xh^2\sim
1/\langle\sigma v\rangle$.

We remark that the scaled Hubble-constant-independent relic abundance of the DM species $X$
behaves as
$$\Omega_X\,h^2\sim {m_X^4}x^{-3}{Y_\infty}\quad(x\to\infty).$$
Also, for the case of standard cosmology $Y_\infty={\rm constant}$ [see (\ref{e62})],
\begin{equation}
\Omega_{X}h^2\sim x^{-3}\to0\quad(x\to\infty).
\label{e92}
\end{equation}

For dilaton cosmology [see (\ref{e3})] one has a modified law
\begin{equation}
\Omega_{X}h^2\sim x^{-3-|\Phi_0|}\to0\quad(x\to\infty).
\label{e93}
\end{equation}

Finally we remark that for the dilaton case $\Phi_0>0$,  the string coupling would increase for
large times, and the theory would become strongly coupled and thus intractable.
Nonperturbative string corrections would need to be incorporated. Nevertheless,
the formal solution for $Y(x)$ behaves asymptotically as $Y(x)\sim x^{|\Phi_0|
}$. This would still imply an asymptotically vanishing relic abundance provided
that $\Phi_0<3$ because we have $\Omega_X h^2
\sim x^{-3+|\Phi_0|}\to0$ as $x\to\infty$.

\bigskip
{\bf (ii) Stochastic D-Particle-Foam Cosmology}
\bigskip

It is known that in the background of  D-particle space-time foam (for
constant dilatons), the Boltzmann equation for the number density $n_X$ of the
DM species $X$ assumes the form \cite{R8}
\begin{equation}
\frac{d}{dt}n_X+3Hn_X=\Gamma_{\rm D-foam}(t)n_X+C[n],
\label{e94}
\end{equation}
where $\Gamma_{\rm D-foam}(t)=2Ha^4m_X\ll\Delta^2\gg\frac{g_s^2}{M_s^2}T\left(9+2m_X/T\right)$.
The notation and conventions here are those of \cite{R8}, where $C[n]=\langle
\sigma v\rangle\big[(n_X^{\rm eq})^2-(n_X)^2\big]$ is the standard nonlinear
interaction term and $n_X^{\rm eq}$ is the thermal equilibrium number density of $X$.
As discussed in \cite{R8} and reviewed in the text [see Eq.~(\ref{e26})], the
recoil fluctuations of the D-foam $\ll\Delta^2\gg$ averaged over populations of
D-particle defects, have the scaling $\ll\Delta^2\gg\sim\langle\Delta^2\rangle_0
a^{-3}$.

To this end, we use the cooling law (\ref{e27}) and ignore the nonlinear
interaction term $C[n]$. From (\ref{e94}) and (\ref{e26}), for the regime of temperatures $m_X
\gg T$, the energy density $\rho_X=m_X n_X$ then satisfies the continuity equation
\begin{equation}
\frac{d}{dt}\rho_X+3H\rho_X={\tilde\Gamma}_{\rm D-foam}(t)H\rho_X,
\label{e95}
\end{equation}
where ${\tilde\Gamma}_{\rm D-foam}(t)=4C_0 \frac{g_s^2}{M_s^2}\frac{m_X^2}{T}\langle\Delta^2\rangle_0$.
For weak foam effects we have $\tilde \Gamma_{\rm D-foam} < 1 $ in the range of temperatures we are interested in; that is from the early universe until today ($T \ge C_0$).
Equivalently, for an expanding universe where ${\dot a}>0$ we have
\begin{equation}
d(\rho_X V)-\rho{\tilde\Gamma}_{\rm D-foam}dV/{3}=0.
\label{e96}
\end{equation}

Equation (\ref{e96}) implies that the thermodynamic interpretation of heavy DM dust in the foam
background is that of a gas with an adiabatic expansion of its volume. During the expansion
entropy is constant, and the effective pressure $p_{{\rm eff}-X}$ of the
gas is {\it negative} (indicating cosmological instabilities):
\begin{equation}
p_{{\rm eff}-X}=-\rho{\tilde\Gamma}_{\rm D-foam}/3.
\label{e97}
\end{equation}
Note that $p_{{\rm eff}-X}$ has a nontrivial dependence on the temperature.

From the cooling law (\ref{e27}), we may write (see [\ref{e95}]) ${\tilde \Gamma} \equiv {\tilde \gamma} a $, where ${\tilde \gamma}$ is a  constant much less than one. Hence, the scaling of the dust energy density, due to its
interaction with the D-foam, is easily obtained from (\ref{e96}) to be (in units of  $a_0$):
\begin{equation}\label{e96b}
\rho_X  \sim a^{-3} \, e^{{\tilde \gamma} \, \int _1^a   da }
 \sim T^3 e^{{\tilde \gamma} \, \big({C_0}/T - 1\big) }
\end{equation}

To find the entropy function that remains constant it is essential to consider the total energy density $\rho$, including relativistic degrees of freedom, and not only $\rho_X$.
In a similar spirit to the dilaton case, the relativistic degrees of freedom are insensitive to the heavy D-foam source effects; in this sense they satisfy an equation
of the form (\ref{e78})
by themselves with equation of state $p =\rho/3$, which can be added to (\ref{e96})  to give the equation
\begin{equation}\label{e96c}
 d((\rho^{\rm rad} + \rho_X) V) + \big(p^{\rm rad} + p_{{\rm eff}-X}  \big) dV \equiv   d(\rho V) + p_{\rm eff} dV = 0.
 \end{equation}
Equation (\ref{e96c}) is the analog of the continuity equation in the case of D foam. We stress that in (\ref{e96c}) $\rho $ and $p_{\rm eff}$ refer to the \emph{total }energy density and pressure, including relativistic degrees of freedom and D-foam background effects.

Taking into account that $\rho$ is a function of $T$, we can formally replace the right
side of (\ref{e96c}) by $Td\mathcal{S}$ to determine the (constant) entropy function $\mathcal{S}$ (ignoring  chemical potential terms, a valid assumption for weak D-foam);
only at the very end of the computation will we set $d\mathcal{S}$ to zero. We
then have
\begin{equation}
d\mathcal{S}=\frac{V}{T}\frac{d\rho}{dT}dT+\frac{\rho+p_{\rm eff}}{T}dV.
\label{e98}
\end{equation}
$\mathcal{S}$ is considered to be a smooth function of $T$ and $V$, which are
treated as independent variables. From the requirement (\ref{e84}) we deduce
the condition
\begin{equation}
-\frac{\rho+p_{\rm eff}}{T^2}+\frac{1}{T}\frac{dp_{\rm eff}}{dT}=0.
\label{e99}
\end{equation}

We then see immediately from (\ref{e98}) and (\ref{e99})  that
$$d\mathcal{S}=d\left[\frac{\rho+p_{\rm eff}}{T}V\right]-\frac{V}{T}\frac{dp_{\rm
eff}}{dT}dT+\frac{\rho+p_{\rm eff}}{T^2}VdT = d\left[\frac{\rho+p_{\rm eff}}{T}V\right],$$
which upon setting $dS=0$ implies the constancy of the effective entropy
function in the co-moving volume $V$:
\begin{equation}\label{e99b} \mathcal{S}=S_{\rm eff}=\frac{\rho+p_{\rm eff}}{T}V = {\rm constant}.\end{equation}
Note that we have used (\ref{e96b}) and the cooling law (\ref{e27}); $\rho$ and $p_{\rm eff}$  refer to the
total energy density and pressure including
relativistic components, which is essential for consistency.
As in the previous cases, the relativistic degrees of freedom  dominate the entropy.
The entropy density $s$ associated with $\mathcal{S}$ is given by an expression similar in form to that in standard cosmology:
\begin{equation}
s_{\rm eff~D-foam}={\rho + p_{\rm eff}}/T
\label{e100}
\end{equation}
and scales with the temperature as $T^3$. Hence, $s$ can be treated as a fiducial quantity to
define $Y(x)$ just as in the dilaton cosmology case (i) above.

Notice also that for the case of dust in dilaton cosmology, the effective
entropy function (\ref{e88}) is reproduced upon replacing the source ${\tilde
\Gamma}_{\rm D-foam}$ by the corresponding source of the running dilaton
(\ref{e3}) cosmology ${\tilde\Gamma}_{\rm running~dil}=-|\Phi_0|$. (With our
definitions we have $\Gamma_{\rm running~dil}={\dot\Phi}=-|\Phi_0|H\equiv
{\tilde\Gamma}_{\rm running~dil}H$.)

In the paper we considered the combined source case, where the foam appears
together with a nontrivial running dilaton of the form (\ref{e3}). The string
coupling $g_s=e^\Phi$ exhibits a nontrivial scaling with the scale factor and
also with temperature. The combined source is taken to be the {\it algebraic
sum} of the respective two source terms; that is,
\begin{equation}
\Gamma_{\rm total}\equiv{\tilde\Gamma}_{\rm total}H=\left[-|\Phi_0|+4
\frac{g_{s0}^2 \, m_X^2}{M_s^2}\langle\Delta^2\rangle_0
\left(\frac{C_0}{T}\right)^{1 - 2|\Phi_0|}
\right]H
\label{e101}
\end{equation}
in the asymptotic region $m_X\gg T$ of interest, where we have assumed the
cooling law (\ref{e27}).


\begin{thebibliography}{10}

\bibitem{R1} C.~M.~Bender and S.~Sarkar,
J.~Math.~Phys.~{\bf 53}, 103509 (2012), arXiv:1203.1822 [hep-th].

\bibitem{R2} C.~M.~Bender and S.~A.~Orszag, \textit{Advanced Mathematical
Methods for Scientists and Engineers} (McGraw Hill, New York, 1978).

\bibitem{R3} E.~W.~Kolb and M.~S.~Turner, \textit{The Early Universe} (Westview,
Boulder, 1994); S.~Dodelson, \textit{Modern Cosmology} (Academic, New York,
2003);  R.~J.~Scherrer and M.~S.~Turner, Phys.~Rev.~D {\bf 33}, 1585
(1986).

\bibitem{R4} M.~Gasperini, \textit{Elements of String Cosmology} (Cambridge
University Press, Cambridge, 2007).


\bibitem{R5} A.~B.~Lahanas, N.~E.~Mavromatos, and D.~V.~Nanopoulos,
PMC Phys.\ A {\bf 1}, 2 (2007) {[}hep-ph/0608153{]}.


\bibitem{R6} A.~B.~Lahanas, N.~E.~Mavromatos, and D.~V.~Nanopoulos,
Phys.~Lett.~B {\bf 649}, 83 (2007) {[}hep-ph/0612152{]};
A.~B.~Lahanas,
Phys.~Rev.~D {\bf 83}, 103523 (2011), arXiv:1102.4277 {[}hep-ph{]};
A.~B.~Lahanas and V.~C.~Spanos,
JHEP {\bf 1206}, 089 (2012), arXiv:1201.2601 [hep-ph].

\bibitem{wmap} E.~Komatsu {\it et al.} [WMAP Collaboration],
Astrophys.~J.~Suppl.~{\bf 192}, 18 (2011), arXiv:1001.4538 [astro-ph.CO].


\bibitem{R7} B.~Dutta, A.~Gurrola, T.~Kamon, A.~Krislock, A.~B.~Lahanas,
N.~E.~Mavromatos, and D.~V.~Nanopoulos,
Phys.~Rev.~D {\bf 79}, 055002 (2009), arXiv:0808.1372 {[}hep-ph{]}.

\bibitem{R8} N.~E.~Mavromatos, V.~A.~Mitsou, S.~Sarkar, and A.~Vergou,
Eur.~Phys.~J.~C {\bf 72}, 1956 (2012), arXiv:1012.4094 {[}hep-ph{]};
N.~E.~Mavromatos, S.~Sarkar, and A.~Vergou,
Phys.~Lett.~B {\bf 696}, 300 (2011), arXiv:1009.2880 {[}hep-th{]}.

\bibitem{R9} C.~M.~Bender, K.~A.~Milton, S.~S.~Pinsky, and L.~M.~Simmons, Jr.,
J.~Math.~Phys.~{\bf 30}, 1447 (1989).

\bibitem{R10} K.~Huang, \textit{Introduction to Statistical Physics}
(Taylor and Francis, London, 2001).

\bibitem{R11} R.~M.~Corless, G.~H.~Gonnet, D.~E.~G.~Hare, D.~J.~Jeffrey,
and D.~E.~Knuth, Advances in Computational Mathematics {\bf 5}, 329 (1996).


\bibitem{R12} S.~Basilakos, N.~E.~Mavromatos, V.~A.~Mitsou, and M.~Plionis,
Astroparticle Physics {\bf 36}, 7 (2012),
DOI information: 10.1016/j.astropartphys.2012.04.007,
arXiv:1107.3532 [astro-ph.CO].

\bibitem{yusaf} N.~E.~Mavromatos, M.~Sakellariadou, and M.~F.~Yusaf,
arXiv:1211.1726 [hep-th].

\end{thebibliography}
\end{document}